\documentstyle[12pt,epsf]{article}

\begin{document}

\begin{center}{{\bf
On the Deconfinement Phase Transition in 
Hot Gauge Theories with Dynamical Matter Fields}
\footnote{Work supported by Bundesministerium f\"ur Wissenschaft,
Forschung und Kunst of Austria.}  \\
\vglue 1.0cm

O.~Borisenko, M.~Faber  \\
\vglue 0.2cm
\baselineskip=14pt
{\it Institut f\"ur Kernphysik,  Technische Universit\"at Wien,} \\
\baselineskip=14pt
{\it Wiedner Hauptstr. 8-10, A-1040 Vienna, Austria} \\
\vglue 0.4cm

G.~Zinovjev \\
\vglue 0.2cm
\baselineskip=14pt
{\it Institute for Theoretical Physics, National Academy of Sciences
of Ukraine, Kiev 252143, Ukraine}}\\
\vglue 0.4cm

\end{center}
\vglue 0.6cm

\begin{abstract}
The phase structure of hot gauge theories with dynamical matter fields
is reexamined in the canonical ensemble with respect to triality. 
Since this ensemble implies a projection to the zero triality sector of the 
theory we introduce a proper quantity which is able to reveal a critical 
behaviour of the theory with fundamental quarks. We discuss the properties 
of both the chromoelectric and chromomagnetic sectors of the theory and show 
while electric charges carrying a unit of $Z(N_c)$ charge are screened at 
high temperatures by dynamical matter loops, this is not the case for the 
$Z(N_c)$ magnetic flux. An order parameter is constructed to probe the 
realization of {\it local} discrete $Z(N_c)$ symmetry in the magnetic sector. 
We argue it can be used to detect a deconfinement phase being defined in
terms of the screening mechanism as a phase of unscreened $Z(N_c)$ flux. 
It may be detectable at long range via the Aharonov-Bohm effect. We discuss 
the possible phase structure of QCD in this approach.
\end{abstract} 

\newpage
 
\section{Introduction}

A new round of considerable interest in discrete global and local gauge 
symmetries started when it was conjectured that the discrete center $Z(N_c)$
of an underlying gauge group $G$ can be of crucial importance for quark
confinement \cite{hooft,mack1}. Another mile-stone appeared
with understanding the deeply rooted relation between the spontaneous
breaking of the global $Z(N_c)$ symmetry and the deconfinement phase
transition in pure gauge models on the lattice and in continuum \cite{gl}. 
This paper continues the previous studies of hot gauge theories 
with dynamical matter fields and emphasizes the importance of the magnetic
sector in the investigation of their phase structure reliably. We propose a new order 
parameter to test the realization of the discrete $Z(N_c)$ symmetry at finite 
temperature and to measure screening effects in different regions of 
temperature and couplings. The result seems equally applicable to any $Z(N_c)$
and $SU(N_c)$ gauge models coupled to Higgs and/or fermion fields carrying
$Z(N_c)$ charge, i.e. which are nontrivial on the center group.

A brief overview of commonly recognized results of finite temperature QCD
analysis occurs quite pertinent for an introduction to the subject. So then,
gluon fields are strictly periodic in time, while
quarks are antiperiodic with a period given by the inverse temperature.
Pure gauge theory has an exact $Z(N_c)$ global symmetry.
The gauge invariant operator, the Polyakov loop (PL),
$L_{\vec{x}} = \frac{1}{N_c} Tr \prod_{t=1}^{N_t} U_0(\vec{x},t)$
transforms under $Z(N_c)$ global transformations as
\begin{equation}
L_{\vec{x}} \rightarrow Z L_{\vec{x}}, \
Z = \exp [\frac{2\pi i}{N_c}n], \ n = 0,...,N_c-1.
\label{PLtr}
\end{equation}
\noindent
The PL can be used as an order parameter to test $Z(N_c)$ symmetry in
pure gauge theory. The expectation value of the PL is interpreted as the free
energy of a probe quark $F_q$ immersed in a pure gluonic bath
\begin{equation}
<L_{\vec{x}}> = \exp (-\frac{1}{T}F_q).
\label{probeq}
\end{equation}
\noindent
Unbroken $Z(N_c)$ symmetry implies $<L>=0$ and $F_q=\infty$.
When the global $Z(N_c)$ symmetry is spontaneously broken,
$<L>$ develops a non-zero value giving a finite magnitude to $F_q$, i.e. 
it costs a finite energy only to create a single quark in the gluonic bath.
However, in \cite{gauss} it was discussed that in a system with a finite
ultraviolet (UV) cutoff the free energy should not diverge. Usually, Monte
Carlo (MC) and analytical calculations are also performed with periodic 
boundary conditions (PBC) in space directions. It has been shown, however, 
that the Gauss' law and this PBC in space are inconsistent unless the sum of 
quark and gluon colour charges vanishes, what looks controversial since they 
have different triality. In fact, it was concluded that for space PBC the 
expectation value of the PL is not the free energy of heavy quark.

The second trouble appears if we realize that in the spontaneously
broken phase $<L>$ may pick up $N_c$ different values corresponding
to $N_c$ equivalent minima of the free energy. Thus, $L$ can be negative
or even complex and Eq.(\ref{probeq}) tells us that the free energy could be
some complex number. This gives rise to doubts that Eq.(\ref{probeq}) 
has the proper physical interpretation. When dynamical quarks
are included the picture becomes more complicated and new troubles appear.
The fermion determinant generates loops winding around the lattice a number of
times which is not a multiple of $N_c$. Such loops present a propagation of
single quarks and transform non trivially under $Z(N_c)$. This means that
dynamical quarks break $Z(N_c)$ symmetry explicitly and screen sources of
heavy quarks at any temperature. The expectation value of the PL prefers
the phase with arg$L=0$ which provides the minimum of the free energy.
Other $Z(N_c)$ phases with 
arg$L = \frac{2\pi i k}{N_c}, k=1,\ldots,N_c-1$ become metastable. 
They possess, however, such unphysical properties as a complex
free energy or entropy \cite{kogan1}. Recently, it has been discovered that
chiral symmetry is not restored in $Z(N_c)$ phases \cite{chr},
what led to another reexamination of degenerate $Z(N_c)$ phases and interfaces
between them in pure gauge theory \cite{hanson} and one of the 
conclusions is that all $Z(N_c)$ phases correspond to the same physical state 
and the interfaces are unphysical. The operator, which we introduce here may 
test directly the $Z(N_c)$ phases and hence can be used to clarify this 
important problem in the description of finite temperature QCD.

The main motivation of our study comes, however, from the following.
It became almost a dogma to consider that the deconfinement 
phase transition is related to the appearance of nonzero triality 
above the critical point. A number of `order parameters' were 
constructed aiming to detect a phase transition and nonzero 
triality states \cite{md,mo}. To our knowledge all of them failed to 
display a phase transition and to test any nontrivial triality state.  
The most conclusive results here have been
obtained in $Z(N_c)$ gauge theories with Higgs fields \cite{mo}.
Since it had been known that at zero temperature this system
has two phases, a confining/screening phase and a deconfining phase
\cite{mm} separated by a critical line and since such
a critical line was not found at finite temperature it was
concluded that presumably at finite temperature the critical behaviour 
is not present at all. Moreover, it has been claimed that
there is no smooth relation of finite and zero temperature theory
because no phase has been found at arbitrary small (but finite) 
temperature which could correspond to a deconfining phase at zero
temperature in the same interval of bare coupling values. By now it became
almost usual to define deconfinement as a `crossover'
but not a genuine phase transition in full theory
and to study the `phase transition' related
to chiral symmetry restoration at light quark masses \cite{detar}. 
The question how this phase transition is related to the deconfining crossover
at larger quark masses has not been answered positively yet. 

Two points are to be discussed here. First of all,  
the deconfined phase was identified with the phase where
nonzero triality should exist and order parameters have been 
constructed to detect such `free triality states'. 
The correctness of such a conjecture is not obvious a priori. 
In the context of this model, 
we demonstrate here that there might be a phase 
transition which, however, is not related to the liberation of triality
but rather to different screening mechanisms of triality
and introduce an operator which tests these different mechanisms.  
Second, the previous emphasizes have been put on
the possible realization of the $Z(N_c)$ global symmetry in
the electric sector of the theory as explained above. The magnetic
sector has been untouched from this point of view.
We intend to overcome this deficiency in the present paper
putting the problem on a rigorous basis and suggesting to study
local $Z(N_c)$ symmetry in this sector.

Certainly, we need an appropriate framework to study the problem.
The grand canonical ensemble (GCE) with respect to
triality is obviously not a proper tool since 
the $Z(N_c)$ symmetry is explicitly broken in this ensemble.
In our previous papers we introduced an ensemble canonical
with respect to triality \cite{trl} (see also \cite{versus}) to
analyze the QCD phase structure \cite{preprint,epce}. 
Since the $Z(N_c)$ symmetry is not explicitly broken in the 
canonical ensemble (CE) it is more suitable to study
the QCD phase structure, especially the chromoelectric sector,
therefore we shall use this formulation in the present paper as well.
Of course, in general we expect in the thermodynamic limit
the same phase structure both in the GCE and in the CE.

The paper is organized as follows.

In Section 2 we introduce the $Z(N_c)$ invariant quantities, 
$A_t$ and $A_s$, which can probe a unit of $Z(N_c)$ charge or flux 
and discuss its properties. These quantities give 
an opportunity to treat the chromoelectric and chromomagnetic sectors of 
the theory on the same footing applying them for the investigation
of the ground state. In Section 3 we discuss in some details the
general phase structure of gauge theories at zero and finite 
temperatures and outline a realization of local $Z(N_c)$ symmetry 
in the chromomagnetic sector of the theory different from the usually
discussed chromoelectric sector.
We define the deconfinement phase by
analogy with the zero temperature deconfinement phase, i.e. as a 
phase where the kinetic screening due to the gluon-gluon interaction
is stronger than the dynamic screening due to dynamical
matter loops and propose an order parameter to test the corresponding
phase transition. Further, we explain how to investigate the problem
of domain walls in QCD and suggest a possible resolution. 
To check our predictions we study in the Section 4 a model of 
$Z(N_c)$ gauge fields coupled to discrete Higgs fields.
We calculate the order parameters $A_t$ and $A_s$ in different
sectors and for various regimes of coupling constants and temperature.
Finally we summarize our results and outline perspectives  
for the application of our results for $Z(N_c)$ and $SU(N_c)$ 
models with dynamical quarks.

\section{ An order parameter for $Z(N_c)$ charges and domain walls}

In this section we construct an order parameter at finite temperature
which may probe a nontrivial charge and/or flux in the Debye screened 
phase and can be represented by a $Z(N_c)$ invariant operator in a finite
volume. Such a quantity was introduced some years ago in the
context of gauge theories with nontrivial discrete abelian center
at zero temperature \cite{kraus2}. We will discuss that this 
quantity is able to detect a nontrivial phase of some state in a
region enclosed by a surface $\Sigma$ 
\begin{equation}
A(\Sigma ,C) = \frac{<F(\Sigma) W(C)>}{<F(\Sigma)><W(C)>} \ ,
\label{OPPK}
\end{equation}
\noindent
with the limits $\Sigma , C \rightarrow \infty$ taken.
$W(C)$ is a space-time Wilson loop nontrivial on the $Z(N_c)$ subgroup.
$\Sigma$ may be thought as a set of plaquettes in a given time-slice
whose dual plaquettes $\Sigma^{*}$ form a closed two-dimensional
surface for this time slice. Further, $\Sigma$ has to be chosen
in such a way that it encloses at the considered slice a time-like
line of the loop $C$. As will be discussed below, the operator
$F(\Sigma)$ frustrates the plaquettes of $\Sigma$ by the so-called
``singular gauge transformations'' with a non-trivial element of
$Z(N)$. 

The action of this order parameter is based on
the Aharonov-Bohm effect: despite the absence of electric
fields a singular potential influences particles at arbitrary
long distances giving them a nontrivial phase during winding
around the solenoid. In our case a $Z(N_c)$ charge may be viewed
as a kind of such a singular solenoid placed at some space position.
To get this configuration one should take the Wilson loop $W(C)$ and consider
the limit $C \rightarrow \infty$. Since the Wilson loop is a source of
$Z(N_c)$ charge one can probe it via the ``device'' $F(\Sigma)$. 
A nontrivial $Z(N_c)$ charges may be detected only in the case when the 
fermionic screening is suppressed relatively to the Debye screening
\begin{equation}
\lim_{\Sigma , C \rightarrow \infty} A(\Sigma , C) =
\exp \left [\frac{2\pi i}{N_c} K(\Sigma , C) \right ],
\label{znob}
\end{equation}
\noindent
where $K(\Sigma , C)$ is the linking number of the surface 
$\Sigma$ and the loop $C$. If the triality charge is totally screened 
by fermions one gets
$\lim_{\Sigma , C \rightarrow \infty} A(\Sigma , C) = 1$.

Let us discuss the implementation of this order parameter 
for gauge theories at finite temperature. 
The following discussion is equally applicable for any gauge
theory if it admits an introduction of the PL which transforms 
nontrivially under the discrete center of the gauge group, 
i.e. $SU(N_c)$ or $Z(N_c)$. We consider the lattice gauge model
with a standard Wilson action given by
\begin{equation}
S_W=\lambda \sum_pTrU_p \ ,
\label{Waction}
\end{equation}
\noindent
where $U_p$ is a product of gauge link matrices around a plaquette 
$p$. Then, since the fundamental PL is a source of $Z(N_c)$ charge
at finite temperature one can insert the correlation function
of PLs into (\ref{OPPK}) instead of a Wilson loop to get
\begin{equation}
A_t(\Sigma , R) = \frac{<F(\Sigma) L_0 L_R>}{<F(\Sigma)><L_0 L_R>} \ .
\label{ZNOP}
\end{equation}
\noindent
The operator $F$ which frustrates the plaquettes of $\Sigma$
is defined by 
\begin{equation}
F(\Sigma) = \exp 
\left [ \lambda \sum_{p\in \Sigma}(Z-1)TrU_p \right ] \ ,
\label{Fop}
\end{equation}
\noindent
where $Z$ is a non-trivial element of $Z(N_c)$. One can then 
define the ``frustrated action'' in the following way
\begin{equation}
S_W \rightarrow S_F = 
\lambda \sum_{p\notin \Sigma}TrU_p + \lambda \sum_{p\in \Sigma}ZTrU_p \ .
\label{Faction}
\end{equation}
\noindent
A schematic diagram of the surface $\Sigma$ in space is shown in
Fig.1 for the case of a $(2+1)$-dimensional theory. The two Polyakov
loops $L_0$ and $L_R$ are closed in time direction. The frustrated 
plaquettes $\Sigma$ are drawn, they enclose the loop $L_R$.
The links dual to the plaquettes of $\Sigma$ form a closed line
$\Sigma^{*}$ surrounding $L_R$. They are shown by a dotted line.
The frustration of the plaquettes of $\Sigma$ can also be
achieved by multiplying the time-like links enclosed by $\Sigma$
with $Z$. Bold lines represent these links. 

The canonical partition function reads
\begin{equation}
Z = \frac{1}{N_c} \sum_{k=1}^{N_c} \int \prod_l dU_l \prod_{x,i}
d\bar{\Psi}_x^i d\Psi_x^i  \; e^{-S_W - S_q(k)} \ .
\label{CEpf}
\end{equation}
\noindent
$S_q$ is the standard quark action, where we have to substitute
$U_0\rightarrow \exp [\frac{2\pi ik}{N_c}]U_0$ for one time slice in 
the CE \cite{trl}. Then, $\sum_k$ projects to zero triality states.
We define a new partition function in the CE with 
the action $S_F$ (\ref{Faction}) as 
\begin{equation}
Z_F = \frac{1}{N_c} \sum_{k=1}^{N_c} \int \prod_l dU_l \prod_{x,i}
d\bar{\Psi}_x^i d\Psi_x^i  \; e^{-S_F - S_q(k)} \ .
\label{Fpf}
\end{equation}
\noindent
In what follows we refer to the expectation values $<\cdots>_F$
calculated in the ensemble defined by (\ref{Fpf}) as those  
obtained in the $F$-ensemble to distinguish them from 
the corresponding values $<\cdots>_0$ 
in the standard ensemble.
It is easily seen that with such a definition $A_t$ 
in (\ref{ZNOP}) reads
\begin{equation}
A_t(\Sigma , R) = \frac{<L_0 L_R>_F}{<L_0 L_R>_0} \ .
\label{ZNOP1}
\end{equation}
\noindent
\vspace{0.5cm}
\begin{figure}[ht]
\centerline{\epsfxsize=8cm \epsfbox{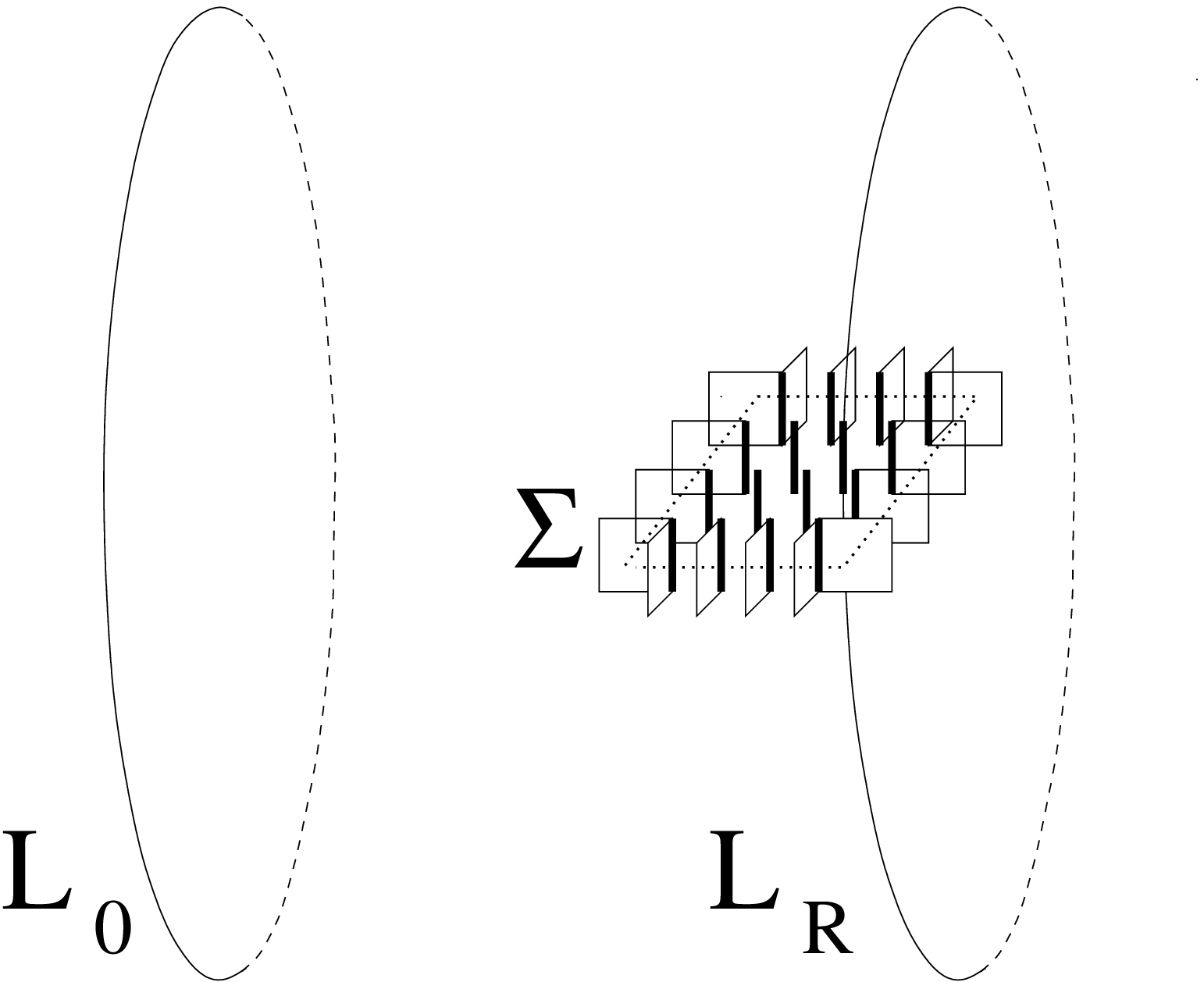}}
\caption{Correlation function of Polyakov loops in the $F$-ensemble
for the case of a (2+1)-dimensional theory. The two Polyakov loops
$L_0$ and $L_R$ extend in time direction. The frustrated plaquettes
are explicitly depicted. Their dual links form a closed (dotted) 
line around $L_R$. Frustration can be achieved by multiplying 
of the bold links with a nontrivial element $Z\in Z(N_c)$.}
\label{fig1}
\end{figure}

To probe a $Z(N_c)$ charge in space one has to take the limit
$R\rightarrow \infty$ where correlations of the ratio (\ref{ZNOP1})
reduce to one PL. Thus, one can probe a unit of
$Z(N_c)$ charge enclosed by the surface $\Sigma$.

A similar $F$-ensemble can be constructed for the spatial Wilson loop.
In this case the order parameter $A$ is essentially the same 
as in zero temperature theory, 
i.e. it is defined in formula (\ref{OPPK}) where we should
insert a pure space-like Wilson loop instead of a time-like one.
In what follows we use the notation $A_s$ to distinguish it
from $A_t$ introduced above. The meaning of the operator in this 
case is however different as we shall discuss in the next section. 

Let us investigate now some simple properties of $A_t$ and $A_s$.
In a pure gauge system the surface $\Sigma$ may be completely gauged
away by a change of variables $U_0(x) \rightarrow Z^*U_0(x)$ 
on the time-like links in the three dimensional volume 
enclosed by the surface $\Sigma$.
It follows that $A=Z\ne 1$ for the pure gauge system since the fundamental 
PL in the origin changes sign, precisely as in the theory at zero 
temperature \cite{kraus2}.  When dynamical fermion (Higgs) fields are 
added, the surface $\Sigma$ cannot be gauged away since 
the corresponding time-like fields $U_0$ will be flipped in 
the fermionic (Higgs) part of the action, so that 
\begin{equation}
S = \lambda \sum_p TrU_p + \alpha \sum_{l\notin \Omega(\Sigma)}Q_l(U_l) 
+ \alpha \sum_{l\in \Omega(\Sigma)}Q_l(Z^*U_l) \ ,
\label{Factiontr}
\end{equation}
\noindent
where $Q_l$ means the density of either the fermion or the Higgs action 
and $\Omega(\Sigma)$ is a volume enclosed by the surface $\Sigma$. 
$\sum_l$ is the sum over links. The same obviously holds for $A_s$
with the corresponding substitution of space-like links instead
of time-like ones.
   
An important interpretation for the influence of $F(\Sigma)$ comes from the 
following observation. $F(\Sigma)$ tries to implement a phase change at the 
spatial surface $\Sigma$. If the global $Z(N_c)$ symmetry is spontaneously 
broken $F(\Sigma)$ produces a stable interface between the volume enclosed 
by $\Sigma$ and the surrounding vacuum. 
The different phases can be detected by the PL
and the order parameter $A_t(\Sigma , C)$. If the symmetry is unbroken
$F(\Sigma)$ cannot produce a stable interface and the value of the PL
is not influenced by $F(\Sigma)$ which is far apart.

The most essential property of $A_t$ and $A_s$ is,
however, that they measure screening effects of dynamical fermion
(Higgs) loops (dynamical screening) together with screening effects of the 
kinetic energy due to the pure gluonic interaction and show which screening is 
stronger at given conditions. The 'Aharonov-Bohm effect' works only 
when screening due to fermion (Higgs) loops is weaker so that it is 
possible to detect a charge (flux) at spatial infinity.

The simplest example where the quantity $A$ can be calculated exactly 
is the Gaussian model for a scalar field coupled to an external field.
We want now to consider this example which may be quite suggestive 
since it illustrates how the operator works; in this case it does not
correspond to any discrete charge but rather shows the competition
between two kinds of screening and measures which one is stronger.
Let us define the partition function of the massive Gaussian model
in an external field $h_x$ as
\begin{equation}
Z_0 = \int_{-\infty}^{\infty} \prod_x d\sigma_x 
\exp \left(- \frac{1}{2}\sum_{x,n} (\sigma_x - \sigma_{x+n})^2
- m^2\sum_x\sigma_x^2 + \sum_x h_x \sigma_x \right).
\label{SGM}
\end{equation}
\noindent
In this case $\Sigma$ consist of links dual to a $(D-1)$-dimensional
surface $\Sigma^{*}$. The operator $F(\Sigma)$ is introduced
on the stack of links $\Sigma$ and has the form
\begin{equation}
F(\Sigma) = \exp \left [ -\frac{1}{2}\sum_{l\in\Sigma}
(\sigma_x+\sigma_{x+n})^2 +\frac{1}{2}\sum_{l\in\Sigma}
(\sigma_x-\sigma_{x+n})^2 \right ] =
\exp \left [ -2\sum_{l\in\Sigma} \sigma_x\sigma_{x+n}  \right ] \ .
\label{GFop}
\end{equation}
\noindent
In what follows we consider a $D$-dimensional periodic lattice with 
period $L$ and number of sites $V=L^D$. The region $\Omega$
enclosed by $\Sigma$ we choose 
to be a $D-1$ torus winding around the lattice. 
Performing now the substitution $\sigma \rightarrow -\sigma$ in
the interior $\Omega$, we can write down
the partition function in the $F$-ensemble as
\begin{equation}
Z_F = \int_{-\infty}^{\infty} \prod_x d\sigma_x 
\exp \left[- \frac{1}{2} \sum_{x,n} (\sigma_x - \sigma_{x+n})^2
- m^2\sum_x\sigma_x^2 
+ (\sum_{x\notin \Omega (\Sigma)} 
- \sum_{x\in \Omega (\Sigma)}) h_x \sigma_x \right].
\label{FGM}
\end{equation}
\noindent

\vspace{0.5cm}
\begin{figure}[ht]
%\centerline{\scalebox{0.5}{\epsfbox{twist.eps}}}
\centerline{\epsfxsize=8cm \epsfbox{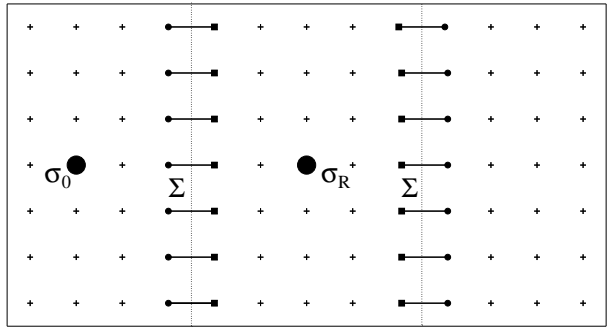}}
\caption{Correlation function of spins in the $F$-ensemble
for the case of a $2D$ theory. The frustrated links $\Sigma$ are 
explicitly depicted. Their dual links form two closed lines 
around $\sigma_R$ and are chosen to wind around the whole lattice.}
\label{fig2}
\end{figure}
After this transformation we have 
\begin{equation}
A = - \frac{<\sigma_0\sigma_R>_F}{<\sigma_0\sigma_R>_0} \ ,
\label{AGM1}
\end{equation}
\noindent
where $<\sigma_0\sigma_R>_0$ and $<\sigma_0\sigma_R>_F$
are the correlation functions calculated in the usual
and the $F$-ensemble, correspondingly.
A diagram of the 
correlation function in the $F$-ensemble is plotted in Fig.2.
On a finite lattice $<\sigma_0\sigma_R>_0$ is known to be
\begin{eqnarray}
<\sigma_0\sigma_R>_0& =& 
\frac{1}{2V}\sum_{k=0}^{L-1}M_k^{-1}\exp [i\frac{2\pi Rk}{L}]
\nonumber   \\
&+& \frac{1}{4V}
\left ( \sum_{k=0}^{L-1}M_k^{-1}h_k \exp [i\frac{2\pi Rk}{L}] \right )
\left ( \sum_{k=0}^{L-1}M_k^{-1}h_k \right ),
\label{G0R}
\end{eqnarray}
\noindent
where
\begin{equation}
h_k = \frac{1}{\sqrt{V}} \sum_{k=0}^{L-1}h_x \exp [i\frac{2\pi xk}{L}] \ 
\label{emf}
\end{equation}
\noindent
and the propagator $M_k^{-1}$ in the momentum space has the well known form
\begin{equation}
M_k^{-1} = (D + m^2 - \sum_n^D\cos \frac{2\pi k_n}{L})^{-1}.
\label{prop}
\end{equation}
\noindent
For $<\sigma_0\sigma_R>_F$ we should change signs 
for $h_x$ in the Fourier transform (\ref{emf}) at corresponding sites.  
Let us consider now the simple  case of a constant external field. We have
from (\ref{emf}) $h_k = \sqrt{V}\delta_{k,0}h$. In the thermodynamic 
limit $V \rightarrow \infty$ we get the following expression
\begin{equation}
A = - \frac{G(R) + h^2/2\left( M_0^{-1}-2f(R) \right)
\left( M_0^{-1}-2f(0) \right)}
{G(R) + h^2/2(M_0^{-1})^2} \ ,
\label{AGM}
\end{equation}
\noindent
where we used
\begin{equation}
G(R) = \int_0^{2\pi}\left( \frac{d\phi}{2\pi} \right)^D 
\frac{e^{i\phi R}}{D + m^2 - \sum_{n=1}^D \cos \phi_n} \ ,
\label{GR}
\end{equation}
\noindent
\begin{equation}
f(R) = \sum_{n=0}^{L_F-1} \int_0^{2\pi}\frac{d\phi}{2\pi} \  
\frac{e^{i\phi (R+n)}}{1 + m^2 - \cos \phi}
\label{fR}
\end{equation}
\noindent
and $M_0 = m^2$. $L_F$ in the last formula denotes the number
of sites with negative sign of $h$. 
In any dimension $G(R)$ goes to zero exponentially as $R$ increases.
In the limit $R$ and $L_F \rightarrow \infty$ one can get the final
expression for $A$
\begin{equation}
A = \frac{m}{\sqrt{2+m^2}} \ . 
\label{AGM2}
\end{equation}
\noindent
One can see from the last formula that for a constant external
field $A$ is always positive. It is interesting to note
that the limit $m\rightarrow 0$ exists even for the one- and two-dimensional
cases and gives $A=0$. The limit $m\rightarrow \infty$ corresponds
to the value $A=+1$. This is the only mass when $A=1$ since the scalar 
field $\sigma$ does not carry a unit of any discrete charge.
As we discussed earlier in Eq. (\ref{znob}), 
for fields carrying a discrete charge 
we would find only discrete values of $A$.

The interpretation of these results is rather transparent.
There is a competition in the correlation function between the 
term $G(R)$ and the second term  of (\ref{G0R}). $G(R)$ comes
from the kinetic energy of the system, decays
exponentially and describes the corresponding
Debye screening. The second term comes from the external field
and describes ``screening'' due to this field.
We think this example shows that $A$ is able to determine 
which screening mechanism is stronger.
In fact, it is not difficult to find a nonconstant external field
such that in the limit $R\rightarrow \infty$ the quantity $G(R)$
goes to zero slower than the second term of the correlation function
and hence $A=-1$. A special example of this type we will mention later.

\section{Screening at zero and finite temperatures and 
the deconfinement phase transition}

We are now ready to proceed to study the phase structure
of gauge models with a nontrivial center. We would like to 
explore the above mentioned properties of $A_t$ 
and $A_s$ to reveal some features of these models.
Let us start with a discussion of pure gauge theory
at zero temperature. 
First, we are interested in $Z(N_c)$ gauge models.
These models are known to exhibit confinement at strong coupling
and deconfinement behaviour at weak coupling. 
This means that the Wilson loop obeys an area law in the 
first case and a perimeter law in the second. The interpretation of 
such a behaviour in terms of the quantity $A$ has been given in 
\cite{kraus2} and amounts essentially to: $A=-1$ and the absence of 
dynamical screening of a $Z(N_c)$ charge at any coupling. 
The Wilson loop in the $F$-ensemble always
changes its sign, though in different ways, in strong and weak coupling
regimes. At weak coupling the ground state of the system consists
of gauge spins flipped inside a volume bounded by $\Sigma$ relatively
to gauge spins outside. A perimeter law for the Wilson loop arises
in this phase due to the kinetic energy of gluons as can be seen
most easily in the Hamiltonian formulation (see \cite{versus} for
a comprehensive discussion of this topic).
If we add Higgs fields to this system dynamical screening appears,
causing a competition with the screening coming from the gluon kinetic
energy. Before we proceed further we need to give
more precise definitions of dynamical screening and of screening
coming from the gluon kinetic energy.
  
Let $W(C)$ be a Wilson loop in $Z(N_c)$ lattice gauge theory 
either at zero or at finite temperature. In the weak coupling
phase of pure gauge theory one may write
\begin{equation}
<W(C)> \sim \exp (-\gamma_{gl} P),
\label{pergauge}
\end{equation}
\noindent
where $P$ means the perimeter of the loop $C$.
We refer to nonzero $\gamma_{gl}$
in the deconfinement phase as coming from the kinetic
energy of the pure gluonic interaction
and attribute the corresponding kinetic screening
to this interaction. Let us imagine now this system coupled 
to dynamical matter fields, either to Higgs or to fermion ones.   
Among other contributions to the Wilson loop the dynamical fields 
generate a perimeter decay of the loop at {\it any} coupling
and/or temperature. In the $C\to\infty$ limit one always finds 
the following behaviour of the loop in the strong coupling limit, 
i.e. $\lambda =0$ in (\ref{CEpf})
\begin{equation}
<W(C)> \sim \exp (-\gamma_{dyn} P).
\label{perhiggs}
\end{equation}
\noindent
We refer to $\gamma_{dyn}$ as coming from the dynamical
screening due to the dynamical Higgs or fermion fields. 

Let us suppose now that the system is in a region of coupling
constants where the pure gauge interaction produces only
area law decay for the Wilson loop, i.e. $\lambda << 1$. 
One has the following formal expansion 
\begin{equation}
<W(C)> \sim K_1\exp (-\gamma_{dyn} P) + K_2\exp (-\alpha_{gl} S) + ... \ ,
\label{conf}
\end{equation}
\noindent
where $\alpha_{gl}$ is a string tension of the pure gauge theory
and $S$ is the minimal area enclosed by the loop $C$.
Dots mean all the (perimeter and area) terms which vanish faster
in the thermodynamic and $C\to\infty$ limits than
the corresponding written terms. From the properties 
of $A$ described in the previous
section it is clear that in the $F$-ensemble the term with area
law decay should change its sign. The expansion might look like 
\begin{equation}
<W(C)> \sim K_1\exp (-\gamma_{dyn} P) - K_2\exp (-\alpha_{gl} S) + ... \ .
\label{Fconf}
\end{equation}
\noindent
Substituting these expansions into (\ref{OPPK})
and taking the limit $C\to\infty$ we find $A=1$ since the area term
coming from the pure gluon interaction vanishes clearly faster
\begin{equation}
A(\Sigma , C) = \frac{<W(C)>_F}{<W(C)>_0} \rightarrow_{C\to\infty} 1. 
\label{Atoconf}
\end{equation}
\noindent
Thus, $A$ takes a trivial value and we interpret this behaviour
as reflecting the following fact: in the confinement region the dynamical
screening is the only one screening present in the system. Actually,
in this region there is not even a real competition.
However, in the pure gauge system there is a critical point
which of course is not a critical point of a full system.
Above that one has instead of (\ref{conf}) 
\begin{equation}
<W(C)> \sim K_3\exp (-\gamma_{dyn} P) + K_4\exp (-\gamma_{gl} P) + ... \ .
\label{deconf}
\end{equation}
\noindent
In the $F$-ensemble one has to change the sign of $K_4$.
There is a competition between the kinetic screening and
the dynamical screening. One gets
\begin{equation}
A(\Sigma , C) = \left\{
\begin{array}{c}
1 , \gamma_{dyn} < \gamma_{gl} , \\
-1, \gamma_{dyn} > \gamma_{gl} .
\end{array}
\right\} \ .
\label{Atodeconf}
\end{equation}
For the second possibility the kinetic screening is stronger and
a $Z(N_c)$ charge can be detected. We interpret this as
a permanent feature of the deconfinement phase.
Simple and maybe well known conclusions are:

1) deconfinement is the phase at a small gauge coupling constant;

2) deconfinement may take place even in the presence of dynamical
screening;

3) $A$ acts equally good for time and space-like 
Wilson loops since there is no actual difference between them
in zero temperature theory. It distinguishes between 
confined/screened phase and deconfined phase.

The first observation is rather important especially in the context
of finite temperature theory. Usually, the deconfinement 
phase is associated with the high temperature region identifying the 
latter with the region of small couplings. In fact,
at finite temperature the original formulation involves two free and
completely independent parameters: the temperature 
and the bare coupling.
The region of small bare coupling needs a priori not to be the phase
of high temperature. In fact, rigorous results on the deconfinement
transition in pure gauge models \cite{rigres} state that as soon as
the temperature is turned on there exists such a coupling below that
the system is in the deconfined phase.

It is also interesting to mention that the equation (\ref{Atodeconf})
predicts an exact equation for the critical line in the theory, namely
\begin{equation}
\gamma_{dyn}(\alpha) = \gamma_{gl}(\lambda ) \ , 
\label{crtline}
\end{equation}
where $\lambda$ and $\alpha$ are the gauge and Higgs (fermion) couplings,
respectively. 

The main question arising from these observations which we would like 
to put here concerns special features of the deconfined
phase at finite temperature relatively to the one at zero T?
We argue that in fact there are no special features, 
breakdown of global and/or local $Z(N)$ symmetry at high temperature 
may be misleading in this respect if one deals with the full theory.
Triality is always a conserved quantity and to reveal the critical 
behaviour one has to look at the screening mechanisms of triality
in different regions of coupling. We therefore 
suggest to define the deconfinement phase in the theory with
dynamical matter fields as the weak coupling phase where the kinetic 
screening due to the gluon interaction is stronger than dynamical
screening. A similar proposition was discussed also in 
\cite{versus}. Technically however the situation at finite 
temperature may differ from the picture discussed above.

Let us discuss now in detail what we expect for
finite temperature theory. We continue considering $Z(N_c)$ 
gauge models in the Euclidean path integral formulation. 
There is no symmetry anymore between electric and magnetic sectors
at finite temperature, in particular,
one can find a drastic difference in the behaviour of the
corresponding Wilson loops, i.e. the spatial Wilson loop
and the thermal (Polyakov) loop. The spatial Wilson loop
behaves essentially as in the zero temperature case. Thus,
the first observation which can be easily deduced from this fact
is that the quantity $A_s$ should be as good an order parameter
at finite temperature as it is at zero temperature because
all formulae above are valid in this case if we substitute
$W(C)$ by the spatial Wilson loop and $A$ by $A_s$. A nontrivial
value of $A_s$ should imply that a unit of $Z(N_c)$ flux
is not screened dynamically and can be detected at long-range.
Therefore, if there is a critical behaviour at zero temperature
it should also be present at finite temperature.
We would like to stress that this observation does not depend
on the ensemble (CE or GCE) used to study the finite 
temperature system (see also next section).

On the other hand, the behaviour of order parameter $A_t$ 
could be qualitatively described as follows.
In the strong coupling phase the correlation function of
PLs calculated in the pure gluonic sector 
goes down exponentially with increasing distance.
The fermionic sector, however generates terms which screen
heavy quarks and lead thus to a constant value of the correlator
even at spatial infinity. This implies $A_t=1$.
In the weak coupling region the pure gluonic sector also
gives finite values for the correlation at spatial infinity
leading in such a manner to the competition with dynamic
screening. In such a situation the direct use of $A_t$ 
as an indicator of a phase transition is not 
possible because both contributions stay finite in
the $R\to\infty$ limit. One may argue, however that $A_t=1$.
To understand qualitatively what one should expect in 
such a situation one can study some simpler models with similar
properties. Such studies were done by us in \cite{spin} on the
example on the $3D$ Ising model where we were able to define
and to calculate an analogue of the quantity $A$ for spin systems
in the presence of an external magnetic field. It was done 
both analytically and in MC simulations. Applying that consideration
to the $Z(2)$ gauge model we conjecture that in the weak coupling
phase the value $A_t=1$ has the following meaning:
The operator $F$ introduces a stable interface in the pure gauge
system. In the weak coupling region the main state is the one
where all the PLs are flipped inside the volume enclosed
by the surface $\Sigma$ relatively to the loops outside of the
volume. $A_t=-1$, and this means that the interfaces are stable. 
Dynamical matter loops try to destroy such interfaces. 
At sufficiently weak coupling the main state has
all loops aligned in one of the $Z(N_c)$ direction 
(in the canonical ensemble). The state with 
the loops flipped inside of $\Sigma$, which could potentially 
lead to a nontrivial value appears to be suppressed by a volume term
whereas the contribution of the state with aligned loops is suppressed
with a surface term. There is a necessary mathematical condition for
$A_t$ to take a nontrivial value: the expansion
for large masses or small Higgs coupling
has to converge in the electric sector and in the thermodynamic limit. 
The consideration in \cite{mo} shows that such an expansion is presumably   
not convergent at any finite temperature. It means in turn that it is
impossible to expand the correlation of the PLs similarly
to the Wilson loop at zero temperatures, e.g. (\ref{conf}) and (\ref{deconf}).
Indeed, if one wants to make such an expansion one would find at the 
first nontrivial order in small Higgs or fermion and large inverse gauge 
couplings the equations for $A_t$ (see next section for details)  
\begin{equation}
A_t(\Sigma , R) = \frac{<F(\Sigma) L_0 L_R>}{<F(\Sigma)><L_0 L_R>}
= \pm 1 \mp VC e^{-4D\lambda},
\label{Atwrongexp}
\end{equation}
\noindent
where $\lambda$ is gauge coupling, $V$ is the volume and the constant 
$C$ is either proportional to the Higgs or to the fermion coupling. 
A similar expansion one can find for the correlator in the $F$-ensemble.
One sees from here that the thermodynamic limit does not exist
already at the first nontrivial order, even for the correlation of
the PLs. One concludes 
that in this situation $A_t$ cannot determine which
screening is stronger. However, it gives an interesting information
on the domain wall structure of the high temperature phase.
The value $A_t=1$ corresponds as explained above to an instability
of the interfaces of pure gauge theory. We argue thus
that the order parameter $A_t$ can be used to resolve a long standing
problem of high temperature QCD, namely the problem of the stability 
of domain walls. In the CE all $Z(N_c)$ directions are degenerate
but in the thermodynamic limit, as the behaviour of $A_t$
shows there is no stable interfaces between regions with different 
orientations of the PL. In the next section we mention 
however a possibility where the interfaces could be stable and $A_t=-1$. 

Let us shortly summarize.
In the confined phase a $Z(N_c)$ charge does not exist in the sense
that quarks are bound in triality zero states, i.e. mesons and
baryons. Whatever small it is, dynamical screening is finite at any
long range. Thus, a 'free' $Z(N_c)$ electric charge cannot be found 
and $A=1$. In the weak coupling phase, quarks may become free and 
kinetic screening appears to play an important role. 
The correlation function is finite
because of this screening which is defined as coming from the pure gluonic
part. It is however rather impossible to distinguish between this
and dynamical screening because both of them lead to a finite value
of the correlation of the PLs at spatial infinity.
The attempt to separate the different types of screening leads to
volume divergences. 
This fact does not imply that the critical behaviour is lost, 
it concerns only purely time-like loops. One can 
use the quantity $A_s$ to measure screening in the magnetic sector, 
i.e. the presence of a unit of the $Z(N_c)$ flux. Such a treatment
should reveal the critical behaviour if it is present at zero temperature.
Indeed, the screening mechanism in the magnetic sector is not much
affected by dynamical matter loops winding around the lattice
in time direction and should be essentially the same 
(in the thermodynamic limit) as at zero temperature. 

In the next section we show how these ideas work in the theory of $Z(2)$ 
gauge spins coupled to Higgs fields.

\section{Application: $A$ in $Z(2)$ gauge theory}

As an application of the proposed approach we would like to
examine here the model of $Z(2)$ gauge spins coupled to discrete 
Higgs fields at finite temperature. 
The canonical partition function of the $Z(2)$ gauge model
is given by the path integral
\begin{equation}
Z = \frac{1}{2} \sum_{\sigma= \pm 1} \sum_{s_l=\pm 1}
 \sum_{z_x=\pm 1} \; e^{S_W + S_H} \ .
\label{z2higgs}
\end{equation}
\noindent
The sum over $\sigma$ is the sum over two n-ality sectors.
We have denoted
\begin{equation}
S_W = \sum_{p_0} \lambda_0 U_{p_0} + \sum_{p_n} \lambda_n U_{p_n} \ ,
\label{z2g}
\end{equation}
\noindent
where $U_{p_n}$ ($U_{p_0}$) is a product of $Z(2)$ gauge link 
variables $s_{\mu}(x)$ around a space(time) plaquette. 
The Higgs action at finite temperature we write down as
\begin{equation}
S_H = S_H^{sp}+S_H^t = \sum_{x,n}h_n z_xs_n(x)z_{x+n}
+ \sum_{x}h_0 z_x \sigma s_0(x) z_{x+0} \ ,
\label{higgsaction}
\end{equation}
\noindent
where both gauge $s_{\mu}(x)$ and Higgs $z_x$ fields 
obey a periodicity condition.
The phase diagram of this theory at zero temperature was well
established long ago \cite{mm,HT0} and was discussed in the previous
section. The phase diagram at finite temperature is rather unclear.
It has been recently claimed that pure
$Z(N_c)$ gauge model may well possess 4 distinct phases in terms
of bare coupling depending on actual values of $\lambda_0$,
$\lambda_n$ and $N_t$ \cite{pureGT}. Ref. \cite{mo} claims
that the coupling of gauge spins to the Higgs field leads 
to a trivial phase structure, the system is in a confining/screening
phase at all couplings and temperatures. The $U(1)$ abelian Higgs model
at nonzero temperature was studied earlier in \cite{ABHT}.
It has been argued that there is no phase
transition in {\it temperature}, neither increasing the temperature
from the confining/screening phase nor from the deconfined phase.
However, there should be a phase transition at finite temperature
in the {\it coupling constant} from the confining/Higgs phase 
to the deconfined phase. 
It was speculated that this phase boundary may disappear
above some finite temperature but the issue remains unsettled.      
We expect a similar scenario to be realized in the $Z(2)$ Higgs model,
i.e. the deconfinement phase transition is a transition in the coupling
constant either at zero or at some finite temperature.

At zero temperature the order parameter $A$ was calculated
for the case of $Z(2)$ gauge spins coupled to 
Higgs fields in Ref.\cite{kraus2}. 
It was shown that $A$ indeed is an order parameter
changing abruptly from $A=1$ (confining/dynamical
screening phase) to $A=-1$ (deconfining phase).

Let us study first $A_t$ in the finite temperature
Higgs model. Shifting the surface $\Sigma$ via $Z(2)$ singular gauge 
transformations to the Higgs part of the action we get
\begin{equation}
A_t(\Sigma , R) = - \ \frac{<L_0 L_R>_F}{<L_0 L_R>_0} \ ,
\label{Athiggs}
\end{equation}
\noindent
where the PL is given by 
\begin{equation}
L_0 = \prod_{t=1}^{N_t}s_0(0,t).
\label{PLhiggs}
\end{equation}
\noindent
The time-like part of the Higgs action reads
\begin{equation}
S_H^t = \sum_{x}h_0(x) z_x \sigma s_0(x) z_{x+0} \ ,
\label{higgstemF}
\end{equation}
\noindent
where
\begin{equation}
h_0\rightarrow h_0(x)=\left\{
\begin{array}{c}
h_0(x\notin \Omega ),  \\
-h_0(x\in \Omega )
\end{array}
\right\}
\label{heff}
\end{equation}
\noindent
and $\Omega$ is as in (\ref{Factiontr}). We would like now to analyze two 
regions of the gauge couplings $\lambda_0$ and $\lambda_n$,
the weak and the strong coupling regions, at arbitrary $N_t$. 
The corresponding expansions are known to be convergent, precisely
as in the zero temperature theory \cite{mm,conv}.

I. $A_t$ in the strong coupling region, $\lambda_0$ and $\lambda_n << 1$.

We want to get an expression for the partition and
the correlation functions which could be used in both ensembles 
and both for $A_t$ and $A_s$. For that purpose we introduce
a space-time dependence in the Higgs couplings and use the general notation
$h_{\mu}$. Following the standard scheme of the strong coupling expansion
we get (up to the third order), e.g. for the partition function the expression
\begin{eqnarray}
Z = C(\lambda) \prod_l \cosh h_{\mu}(x) [ 1 + 
\sum_p \tanh \lambda_{\mu} \prod_{l\in p} \tanh h_{\mu}(x) +  \nonumber  \\
\sum_{p\neq p^{\prime}}^{\prime}\tanh \lambda_{\mu} \tanh \lambda_{\mu^{\prime}}
\prod_{l\in p,p^{\prime}} \tanh h_{\mu}(x) + 
\hat{\sum_{p\neq p^{\prime}}} \tanh \lambda_{\mu} \tanh \lambda_{\mu^{\prime}}
\prod_{l\in p,p^{\prime}}^{\prime} \tanh h_{\mu}(x) ] 
\label{scexpf}
\end{eqnarray}
\noindent
and we introduced the obvious notation $\lambda_{\mu}$. $\sum_p$ means
sum both over time-like and space-like plaquettes, $\sum_p^{\prime}$
means that one has to omit plaquettes which have a link in common
and $\hat{\sum}_p$ means sum over plaquettes which have one link in common.
$\prod_{l\in p,p^{\prime}}^{\prime}$ means that one has to omit that link
from the product which is common for both plaquettes.
At last, 
$$
C(\lambda) = (\cosh \lambda_0)^{N_{p_0}} (\cosh \lambda_n)^{N_{p_n}}, 
$$ 
\noindent
where $N_{p_0}$, $N_{p_n}$ is the number of time(space)-like plaquettes,
correspondingly. The general expansion for the correlation of the PL is
\begin{eqnarray}
<L_0L_R> = \frac{1}{Z} C(\lambda) 
\sum_{s_l=\pm 1} \sum_{z_x=\pm 1} L_0L_R e^{S_H}
[ 1 +  \nonumber   \\
\sum_p \tanh \lambda_{\mu} \ S_p + 
\sum_{p\neq p^{\prime}}\tanh \lambda_{\mu} \tanh \lambda_{\mu^{\prime}} 
\ S_pS_{p^{\prime}}  + O(\lambda^3) ].
\label{crPL}
\end{eqnarray}
\noindent
Now it is straightforward to calculate correlations of the PL in 
both the standard and the $F$-ensemble. Let us suppose that the
frustrated links are in the time slice $t=1$. Since in this case
space couplings $h_n(x)$ are not affected by a singular transformation
one can omit the space-time dependence. It gives up to second order
\begin{equation}
<L_0L_R> = \prod_{t=1}^{N_t}\tanh h_0(0,t)\tanh h_0(R,t)
[1 + 2DN_t \tanh \lambda_0 \tanh^2h_n (1-\tanh^2h_0) ],
\label{crPL1}
\end{equation}
\noindent
where $D$ is the space dimension. Since the linking number of the PL
in the origin and the surface $\Sigma$ is 1, it gives the result
$$
A_t = 1 + O(\lambda^2).
$$
\noindent
In fact, it is obvious that the expression in the square brackets of 
(\ref{crPL}) is an even function of $h_0(t)$ up to the order
$(\tanh \lambda_0)^{L_{\Sigma}}$, where $L_{\Sigma}$ is the radius
of the region $\Omega$. Only on the boundary the corresponding
plaquettes will change signs. Thus,
\begin{equation}
A_t = 1 - N_{\Sigma}C_1(\tanh \lambda_0)^{L_{\Sigma}},
\label{Atsc}
\end{equation}
\noindent
where $N_{\Sigma}$ is the number of frustrated plaquettes
and $C_1$ is independent of $\Sigma$.
It leads in the $\Sigma\to\infty$ limit to the expected result $A_t=1$.

II. $A_t$ in the weak coupling region $\lambda_0 \ , \lambda_n >> 1$.

We fix a static gauge where all gauge spins in time direction $s_0(x)$
are set to 1 except for one time-slice $t=1$ which includes the frustrated
plaquettes on $\Sigma$. The Debye screening comes from the pure 
gluonic action and produces the formula for the correlation of PLs
\begin{equation}
\ln <L_0 L_R> \; \propto \; \exp [-M_D R],
\label{z2cor}
\end{equation}
\noindent
where $M_D$ is the Debye mass. 
Though it may seem at the first glance that the fermionic contribution
to the correlation function is suppressed as $h_0^{2N_t}$ and could 
be neglected relatively to the Debye screening, this is misleading.
The crucial point is that in order to skip this contribution one
should isolate the contribution from the fermionic sector.
To do that, one needs to make an expansion of the correlation function
in small $h_0$. But such an expansion is known to be divergent \cite{mo}.
Thus, there is no direct way to separate the Debye screening from the
fermionic screening in the electric sector. In this situation 
to calculate $A_t$ we should construct a weak coupling expansion
not supposing a small value for the Higgs coupling, take the thermodynamic 
limit and only then we are allowed to make an expansion at small $h_0$.  
This procedure can be easily accomplished at least in the main orders
of the large $\lambda$ expansion if one knows the main state of the gauge
system for time-like links which has to be perturbated.  
In the usual ensemble this is a state with all the links up or down
depending on which $n$-ality sector we consider in the CE. In the 
$F$-ensemble there is a competition between one of these states
and the state where all the time-like spins are flipped inside a volume
bounded by $\Sigma$ relatively to the links outside. The classical action
on these two configurations is, correspondingly
\begin{equation}
S^I = \lambda_0 N_{p_0} + h_0\left( V - 2\Omega (\Sigma) \right),
\label{Iact}
\end{equation}
\noindent
\begin{equation}
S^{II} = \lambda_0 \left( N_{p_0} - 2N_{\Sigma} \right) + h_0 V.
\label{IIact}
\end{equation}
\noindent
Therefore,
\begin{equation}
S^{II} - S^I = -2\lambda_0 N_{\Sigma} + 2h_0 \Omega (\Sigma).
\label{Dact}
\end{equation}
\noindent
On any finite lattice there always exists such a large $\lambda_0$
that the surface term wins and thus the first configuration will represent
the main state. Since the PL is not flipped in this case in the corresponding
triality sector of the CE one finds $A_t = -1$. 
Since, however we have to consider
the limit $\Sigma \to\infty$, which is equivalent to $\Omega\to\infty$ and
$N_{\Sigma}\to\infty$, we get easily convinced that this state gets
metastable when the volume term suppresses the surface term
in this limit. Hence, it is the second state (\ref{IIact}) which gives 
the dominating contribution to the thermodynamic limit. 
Since the PL in the origin flips the sign one finds $A_t = 1$.  

These qualitative arguments can be made more precise if we consider 
an effective model for the PL which can be obtained in the region 
$\lambda_0 >> \lambda_n$, $h_0 >> h_n$.
Summing over space gauge fields and over Higgs fields we come to
the standard effective model for the PL which in this case reads
(we omit all irrelevant constants)
\begin{equation}
Z_{eff} = \sum_{L_x=\pm 1}\exp \left[ \gamma\sum_{x,n}L_xL_{x+n} +
\sum_x \alpha_x L_x \right],
\label{ising}
\end{equation}
\noindent
where $x$ denotes now the sites of the $D$ dimensional lattice.
This is the familiar Ising model in the external magnetic field with 
effective couplings defined as
$$
\tanh \gamma = (\tanh \lambda_0)^{N_t} \ , \
\tanh \alpha = (\tanh h_0)^{N_t}.
$$
\noindent
In the usual ensemble $\alpha_x=\alpha$, while in the $F$-ensemble
one has to change sign in $\alpha$ if $x\in\Omega (\Sigma)$.
We calculated the correlation function in the $F$-ensemble and
the quantity $A_t$ in the Ising model both analytically and in MC 
simulations \cite{spin}. The results found support completely the
qualitative arguments we presented above. The interpretation of such
behaviour we gave in the previous section. 

It is interesting to mention that there is, at least, one obvious
possibility which leads to a nontrivial value $A_t=-1$ even in
the thermodynamic limit. This is the case of an inhomogeneous magnetic 
field in the effective Ising model (\ref{ising}).
For instance, one can take a slowly varying field of the form 
\begin{equation}
\alpha_x = \alpha_0 \cos \frac{2\pi}{L}x,
\label{inhomh}
\end{equation}
\noindent
where $L$ is the linear size of the system. In such a field one
expects that the ground state is again a state with spins flipped
inside a volume enclosed by an appropriate $\Sigma$ and, 
thus $A=-1$ at large values of $\gamma$. At small $\gamma$
$A=1$ as can be easily shown via the high temperature expansion.
This implies that in the model with Ising spins in the external field
(\ref{inhomh}) there could be a phase transition. With respect to the
Higgs coupling it means that $h(x,t)$ may be a constant everywhere
except one time-slice where it has to be of the form (\ref{inhomh}).
Whether such a field could have any physical significance in the context
of gauge models remains however unclear.

We turn now to the order parameter $A_s$. Let $W_s(C)$ be a space-like
Wilson loop in the finite temperature gauge-Higgs model
defined in (\ref{z2higgs}) and $\Sigma_s$ be a two dimensional
surface on a dual lattice, $\Omega_s$ -- the corresponding volume.
Following the above procedure we have 
\begin{equation}
A_s(\Sigma_s , C) = - \ \frac{<W_s(C)>_F}{<W_s(C)>_0} \ .
\label{Ashiggs}
\end{equation}
\noindent
The temporal part of the Higgs action is not affected by the singular
$Z(2)$ gauge transformations whereas for the spatial part
we get in the $F$-ensemble
\begin{equation}
S_H^{sp} = \sum_{x,n}h_n(x) z_x s_n(x) z_{x+n} \ ,
\label{higgsspF}
\end{equation}
\noindent
where $h_n(x)$ is defined similarly to (\ref{heff}). Again, we consider 
the strong and weak coupling phases in the gauge coupling $\lambda$.

I. $A_s$ in the strong coupling region $\lambda_n << 1$.

Using the general expansion (\ref{scexpf}) one gets up to second order
\begin{equation}
<W(C)> = \prod_{l\in C}\tanh h_l
[1 + 2DP_C \tanh \lambda \tanh^2h (1-\tanh^2h) ],
\label{WL1}
\end{equation}
\noindent
where we consider the symmetric case $\lambda_0=\lambda_n$ and $h_0=h_n$,
for simplicity. $P_C$ is the perimeter of the loop $C$. It gives
$$
A_s = 1 - O(\lambda^2).
$$
\noindent
As in the previous case for $A_t$, it is straightforward to see that
\begin{equation}
A_s = 1 - N_{\Sigma_s}C_2(\tanh \lambda )^{L_{\Sigma_s}},
\label{ASsc}
\end{equation}
\noindent
which gives $A_s=1$ in the $\Sigma_s\to\infty$ limit.

II. $A_s$ in the weak coupling region $\lambda_n >> 1$.

The gauge part gives the following contribution to the Wilson loop
for the space dimension $D=3$ 
\begin{equation}
<W(C)> \propto \exp \left[ - 2P_C (e^{-2\lambda})^6 \right].
\label{z2scor}
\end{equation}
\noindent
We again analyze the symmetric case as above.
The fermionic screening is now suppressed as $h_n^{P_C}$, i.e.
\begin{equation}
<W(C)> \propto (\tanh h_n)^{P_C}.
\label{z2sferm}
\end{equation}
\noindent
The crucial point distinguishing this case from the similar expansion in 
the electric sector is that now we are allowed to expand in small $h_n$ 
since there are no loops going around the lattice in space direction which 
could destroy the convergence. It is straightforward to calculate, 
e.g. $<F(\Sigma_s)>$ in leading order of small $h$
\begin{equation}
<F(\Sigma_s)> = \exp \left[ -\delta N_{\Sigma_s} + O(h^6) \right],
\label{z2F}
\end{equation}
\noindent
where $\delta \approx 2h^4 \tanh \lambda$.
If we remove the surface $\Sigma_s$ back to the pure gauge action, then 
the main contribution in the $F$-ensemble comes from configurations
of the gauge fields $s_n(x)$ flipped in the volume $\Omega (\Sigma_s)$
relatively to $s_n(x)$ outside of $\Omega (\Sigma_s)$. Since the linking 
number of the surface $\Sigma_s$ and the Wilson loop $C$ is 1, the Wilson 
loop changes sign in the $F$-ensemble.
The corresponding expansions in large $\lambda$ and in small $h_n$ are
converging. Therefore, in the limit of infinite $\Sigma_s$ and $C$
all the corrections go to zero and we find
\begin{equation}
A_s(\Sigma_s , C) = -1.
\label{z2ZNOP}
\end{equation}
\noindent
This indicates perhaps a deconfinement phase transition in 
the gauge coupling to a phase where the kinetic screening dominates
the dynamical screening. Above the critical point one can detect a unit of
$Z(2)$ flux by the Aharonov-Bohm effect.  
In the main order the critical line is determined as
$$
\exp \left[ - 2 (e^{-2\lambda^c})^6 \right] = \tanh h_n^c \ .
$$
Our last comment concerns the usage of the CE. One can easily check that
$A_s$ exhibits essentially the same behaviour in both
ensembles, at least in the main order of the Higgs coupling which is not
affected by the temporal loops. Thus, both ensembles should lead
to the same phase structure. The use of the CE is more relevant, as we
believe, for the investigation of the $Z(N_c)$ symmetry in the electric sector,
in particular to give a proper description of the metastable states.

\section{Summary}

In this paper we proposed a method which distinguishes
between different screening mechanisms in gauge theories
at finite temperature. This method relies on the action
of the order parameter $A$ used earlier in the context of
zero temperature theory \cite{kraus2} and is based
on the Aharonov-Bohm effect. Various phenomena can be studied
using this approach. The most notable application is
a possibility for describing the deconfinement phase transition
in terms of screening mechanisms. We suggested a general line
for such an application and illustrated it on the example
of $Z(2)$ gauge spins coupled to Higgs fields. Among other 
applications it is worth to mention the following:

1) The quantity $A_t$ can be used to study the problem of 
domain walls and of the stability of the corresponding
interfaces in finite temperature theories with dynamical
matter fields. That this order parameter indeed may show the instability
of the interface in the thermodynamic limit we showed
in \cite{spin} on the example of the $3D$ Ising model. 

2) A very important and long standing problem in the dynamics
of gauge systems (not only for finite temperature ones)
is the difference between the Higgs and the fermionic screening.
In particular, it is needed to explain the non-confining
character of the weak interaction. Since the quantity $A$ 
provides a detailed description of the screening in both systems
this problem can be formulated rigorously in the present approach.

3) We conjecture here that a similar behaviour of $A$ 
can also be found in the model of $Z(N_c)$ gauge spins coupled
to fundamental fermions. The general phase structure of the latter model
may however differ from the Higgs model studied here. Two essential 
points distinguish the fermion model from the Higgs model. First of all,
in the limit of large $\lambda$ one can get a model of a free
fermion gas which does not possess a phase transition. On the other hand,
in the limit of strong coupling one can find a phase transition in the 
fermion sector which is related to chiral symmetry restoration.
It is an interesting opportunity to explore the properties of 
$A$ in both electric and magnetic sectors in order to investigate
a phase structure of this theory and to get a deeper insight into the
connection (if any) between screening mechanisms of triality 
and chiral symmetry restoration. 

4) The application of all previous suggestions and results to 
$SU(N_c)$ gauge models at finite temperature is, in principle 
straightforward though it is technically much more involved.
Summarizing we would like to stress that although a final answer
to the issue of deconfinement in QCD is not available yet, 
certainly the order parameter $A$ should be able to check the scenario
proposed here also in the context of full QCD.
 
We would like to thank J.~Polonyi for interesting discussions
during preparation of this work in Vienna and Debrecen.

\end{document}